\documentclass[12pt,preprint]{aastex}

\shorttitle{The Mass of the Black Hole in NGC 4151}
\shortauthors{Bentz, et al.}

\received{}
\accepted{}

\begin{document}

\title{A Reverberation-Based Mass for the Central Black Hole in NGC~4151}

\author{ Misty~C.~Bentz\altaffilmark{1}, 
         Kelly~D.~Denney\altaffilmark{1},
         Edward~M.~Cackett\altaffilmark{2}, 
         Matthias~Dietrich\altaffilmark{1},
         Jeffrey~K.~J.~Fogel\altaffilmark{3}, 
         Himel~Ghosh\altaffilmark{1},
         Keith~Horne\altaffilmark{2}, 
         Charles~Kuehn\altaffilmark{1,4},
	 Takeo~Minezaki\altaffilmark{5},
         Christopher~A.~Onken\altaffilmark{1,6},
         Bradley~M.~Peterson\altaffilmark{1}, 
         Richard~W.~Pogge\altaffilmark{1},
         Vladimir~I.~Pronik\altaffilmark{7,8},
	 Douglas~O.~Richstone\altaffilmark{3},
	 Sergey~G.~Sergeev\altaffilmark{7,8},
         Marianne~Vestergaard\altaffilmark{9},
         Matthew~G.~Walker\altaffilmark{3}, and
	 Yuzuru~Yoshii\altaffilmark{5,10}}

\altaffiltext{1}{Department of Astronomy, 
		The Ohio State University, 
		140 West 18th Avenue, 
		Columbus, OH 43210; 
		bentz, denney, dietrich, ghosh, peterson, 
                pogge@astronomy.ohio-state.edu}

\altaffiltext{2}{School of Physics and Astronomy, 
		 University of St. Andrews,
		 Fife, KY16 9SS, Scotland, UK;
		 emc14, kdh1@st-and.ac.uk}

\altaffiltext{3}{Department of Astronomy,
		 University of Michigan,
		 Ann Arbor, MI 48109-1090;
		 fogel, dor, mgwalker@umich.edu}

\altaffiltext{4}{Current address:
		 Physics and Astronomy Department,
		 3270 Biomedical Physical Sciences Building,
		 Michigan State University,
		 East Lansing, MI 48824;
		 kuehncha@msu.edu}

\altaffiltext{5}{Institute of Astronomy, 
		 School of Science, 
		 University of Tokyo,
	 	 2-21-1 Osawa, Mitaka, 
		 Tokyo 181-0015, Japan;
		 minezaki, yoshii@mtk.ioa.s.u-tokyo.ac.jp}

\altaffiltext{6}{Current address:
		National Research Council Canada, 
		Herzberg Institute of Astrophysics,
		5071 West Saanich Road,
		Victoria, BC  V9E 2E7, Canada;
		christopher.onken@nrc-cnrc.gc.ca }

\altaffiltext{7}{Crimean Astrophysical Observatory,
		 p/o Nauchny, 98409 Crimea, Ukraine;
		 sergeev, vpronik@crao.crimea.ua}

\altaffiltext{8}{Isaak Newton Institute of Chile,
	         Crimean Branch, Ukraine}

\altaffiltext{9}{Steward Observatory, 
		University of Arizona, 
		933 North Cherry Avenue, 
         	Tucson, AZ 85721; 
		mvestergaard@as.arizona.edu}

\altaffiltext{10}{Research Center for the Early Universe, 
		School of Science,
        	University of Tokyo, 
		7-3-1 Hongo, Bunkyo-ku, 
		Tokyo 113-0033, Japan}

\begin{abstract}

We have undertaken a new ground-based monitoring campaign to improve the
estimates of the mass of the central black hole in NGC~4151.  We measure
the lag time of the broad H$\beta$ line response compared to the optical
continuum at 5100 \AA\ and find a lag of $6.6^{+1.1}_{-0.8}$ days.  We
combine our data with the recent reanalysis of UV emission lines by
\citeauthor{metzroth06} to calculate a weighted mean of the black hole
mass, $M_{\rm BH} = (4.57^{+0.57}_{-0.47}) \times 10^{7} M_{\odot}$.  The
absolute calibration of the black hole mass is based on normalization of
the AGN black hole mass -- stellar velocity dispersion ($M_{\rm BH} -
\sigma_*$) relationship to that of quiescent galaxies by
\citeauthor{onken04} The scatter in the $M_{\rm BH} - \sigma_*$
relationship suggests that reverberation-mapping based mass measurements
are typically uncertain by a factor of 3--4.

\end{abstract}


\section{INTRODUCTION}

Reverberation-mapping (\citealt{blandford82,peterson93}) is the process
of measuring the time lag of variations in the emission lines of an
active galactic nucleus (AGN) relative to variations in the continuum
source due to light travel time effects.  The time lag and the width of
the emission line is directly related to the mass of the central black
hole ($M_{\rm BH}$).  Reverberation-mapping efforts over the past 15
years have led to the compilation of broad-line region (BLR) radius
measurements and black hole mass estimates for 36 active galaxies
(\citealt{peterson04,peterson05}).  \citet{peterson04} undertook a
reanalysis of most of the reverberation-mapping data.  The data were
reanalyzed in a consistent way, paying particular attention to recent
improvements in cross-correlation analysis, measuring the line widths in
the variable part of the emission line, and accounting for the fact that
changes in the mean luminosity of the source could cause emission-line
time lags to vary over longer time scales than the reverberation time
scale \citep{peterson02}.  While the end result was to significantly
decrease the amount of random and systematic errors in a very
inhomogeneous database, there remained a few data sets for individual
objects that were problematic in some way, making it difficult to
determine either time lags and/or line width measurements with high
confidence.  NGC~4151, one of the brightest and best-studied AGN, is one
such source.  Previous monitoring studies of NGC~4151 include those of
\citet{maoz91} and \citet{kaspi96}.

Our motivation for undertaking a new reverberation-mapping campaign on
NGC~4151 is twofold.  Firstly, reverberation mapping is still the only
direct method for estimating central masses in a broad population of
active galaxies, and is more valuable still in that it does not depend
on angular resolution.  Reverberation results also provide the
fundamental calibration for mass estimates based on radius--luminosity
scaling relationships
(\citealt{wandel99,vestergaard02,vestergaard04,mclure02,kollmeier06,vestergaard06}).
Secondly, NGC~4151 is one of the best candidates for measuring the black
hole mass by other means, and therefore affording a direct comparison
between reverberation-based masses and other methods.  Our specific goal
in undertaking this new ground-based monitoring program is to
significantly improve measurement of the lag, or mean response time,
between the continuum and the H$\beta$ emission line, the width of the
H$\beta$ line in the variable part of the spectrum, and the black hole
mass for NGC~4151.

\section{OBSERVATIONS AND DATA REDUCTION}

\subsection{Spectroscopy}

On every clear night between 2005 February 27 and 2005 April 10, we
obtained spectra of the nucleus of NGC~4151 --- a weakly-barred Sab
Seyfert galaxy at $z = 0.00332$ --- with the Boller and Chivens CCD
Spectrograph (CCDS) on the MDM Observatory 1.3-m McGraw-Hill Telescope.
We used a 5\arcsec\ slit with a position angle (PA) of $90 \degr$, and
the typical exposure time was 1200~s. The spectra were reduced in the
usual way using IRAF\footnote{IRAF is distributed by the National
Optical Astronomical Observatory, which is operated by the Association
of Universities for Research in Astronomy, Inc., under cooperative
agreement with the NSF.} and XVista\footnote{XVISTA was originally
developed as Lick Observatory Vista and is now maintained in the public
domain by former Lick graduate students as a service to the community.
It is currently maintained by Jon Holtzman at New Mexico State
University, and is available at http://ganymede.nmsu.edu/holtz/xvista.}.
An extraction width of 17 pixels, corresponding to 12\farcs75, was used
in the spectral reduction.  Each night had between two and four
individual spectra of NGC~4151, which were averaged together to increase
the signal-to-noise (S/N).

Additional spectra were obtained with the Nasmith Spectrograph and
Astro-550 $580 \times 520$ pixel CCD \citep{berezin91} at the Crimean
Astrophysical Observatory (CrAO) 2.6-m Shajn Telescope.  Spectra were
obtained through a 3\arcsec\ slit at PA $= 90 \degr$ and were reduced
in the usual way.  An extraction width of 16 pixels, corresponding to
11\arcsec, was used in the reduction.

Once the spectra were reduced, they were internally calibrated within
each data set.  Because the [\ion{O}{3}] narrow lines are very bright in
NGC~4151, we had saturation problems with the $\lambda 5007$ \AA\ line,
and therefore scaled all the spectra using the $\lambda 4959$ \AA\ line.
This was accomplished by employing the spectral scaling software of
\citet{vanGroningen92}.  A mean spectrum derived from all of the data in
each set was used as the reference spectrum.  The individual spectra
were compared to the reference spectra and scaled so as to minimize the
residuals of the [\ion{O}{3}] $\lambda 4959$ line in the difference
spectrum (produced by subtracting the reference spectrum from the
individual spectrum).  This process results in a common [\ion{O}{3}]
$\lambda 4959$ flux of $3.76 \times 10^{-12}$ ergs s$^{-1}$ cm$^{-2}$.
As the flux ratio of the [\ion{O}{3}] $\lambda 5007$ to [\ion{O}{3}]
$\lambda 4959$ line is 2.94, our measurement of the [\ion{O}{3}]
$\lambda 4959$ flux is consistent with the values of $F$([\ion{O}{3}])
$= 1.575 \times 10^{-11}$ ergs s$^{-1}$ cm$^{-1}$ for the combined
[\ion{O}{3}] flux determined by \citet{kaspi96} and $F$([\ion{O}{3}]
$\lambda 5007$) $= 1.19 \pm 0.06 \times 10^{-11}$ ergs s$^{-1}$
cm$^{-2}$ determined by
\citet{antonucci83}.

The light curves were created by measuring the average continuum flux
between the observed-frame wavelengths 5100--5130~\AA.  To measure the
H$\beta$ line flux, we use observed-frame wavelength windows of
4770--4800~\AA\ and 5100--5130~\AA\ to estimate a linear continuum and
then integrate the flux above the best fit continuum between
4810--4940~\AA.

\subsection{Photometry}

Photometric $V$-band data was obtained with the multicolor imaging
photometer (MIP) at the 2.0-m Multicolor Active Galactic NUclei
Monitoring (MAGNUM) telescope at the Haleakala Observatories in Hawaii
(\citealt{kobayashi98a,kobayashi98b,yoshii02,yoshii03}).  The telescope
was pointed first at NGC~4151 and then at two reference stars with
($\Delta \alpha, \Delta \delta$) = (16\farcm1, 5\farcm7) and
(-11\farcm3, -9\farcm3), after which the telescope was dithered.  The
reference stars were previously determined to be non-variable
\citep{minezaki04}.

The images were reduced with IRAF in the usual way, with an additional
small correction applied for the nonlinearity of the detector.  Aperture
photometry with an aperture diameter of 8\farcs3 was applied to the
nucleus of NGC~4151 and the two reference stars, with sky flux measured
between radii of 5\farcs5--6\farcs9.  The nuclear flux of NGC~4151
relative to the reference star at ($\Delta \alpha, \Delta \delta$) =
(16\farcm1, 5\farcm7) was estimated to be the average of the relative
fluxes for all of the dithering sets.  The reference star flux was
calibrated with respect to photometric standard stars taken from
\citet{landolt92} and \citet{hunt98}.

\subsection{Intercalibration of Light Curves}

Each of these data sets involve different aperture geometries and seeing
conditions, which result in different slit losses and varying amounts of
starlight contamination.  Therefore, they must be intercalibrated to fit
each other before merging them into a final light curve.  The MDM
data set was the largest, and was therefore the anchor to which the CrAO
and MAGNUM data were scaled.  Following the discussion of
\citet{peterson91}, the CrAO and MAGNUM light curves were each compared
to the MDM light curves to identify pairs of observations that were
within one day of each other.  A least-squares analysis was then
performed on the sets of pairs of observations to identify the overall
offsets, the results of the various monitoring geometries, between the
MDM data set and the CrAO and MAGNUM data sets.  After removal of the
offsets, the CrAO and MAGNUM light curves were merged with the MDM light
curves.

The full light curves are given in Table~1.  For our analysis, we binned
all observations that occurred within a 0.5 day window.  The statistical
properties of the binned light curves are given in Table~2.  Column (1)
gives the spectral feature and column (2) gives the number of
measurements in the light curve.  Columns (3) and (4) are the sampling
intervals between data points, measured as the mean and median,
respectively.  Column (5) is the mean flux and standard deviation of the
light curve, not including a correction for host-galaxy starlight,
which, following the methods of \citet{bentz06} for this slit geometry,
is measured to be $F_{\rm gal} ({\rm 5100 \AA}) = (2.11 \pm 0.18) \times
10^{-14}$~ergs~s$^{-1}$~cm$^{-2}$~\AA$^{-1}$.  Column (6) gives the mean
fractional error, which is based on the comparison of observations that
are closely spaced in time.  The excess variance in column (7) is
computed as

\begin{equation}
F_{\rm var} = \frac{\sqrt{\sigma^2 - \delta^2}}{\langle f \rangle}
\end{equation}

\noindent where $\sigma^2$ is the variance of the fluxes, $\delta^2$ is 
their mean-square uncertainty, and $\langle f \rangle$ is the mean of
the observed fluxes.  Finally, column (8) is the ratio of the maximum to
the minimum flux for each lightcurve.

\section{DATA ANALYSIS}

\subsection{Time Series Analysis}

To measure the lag between the continuum and the variable part of the
H$\beta$ line, we cross-correlate the H$\beta$ emission-line light curve
with the continuum light curve measured at 5100 \AA.  We use the
Interpolation Correlation Function (ICF) method as described by
\citet{white94}.  The light curves and cross-correlation functions are
shown in Fig.~1.

To better quantify the uncertainties in the time delay measurement, we
employ the model-independent Monte Carlo FR/RSS method, as described by
\citet{peterson98}, including modifications described by
\citet{peterson04}.  For each single realization of the method,
``random subset sampling'' (RSS) is employed in which a light curve with
$N$ data points is randomly sampled $N$ times without regard to the
previous selection of each point.  A data point that is selected $M$
times will have its uncertainty reduced by a factor $M^{1/2}$.  ``Flux
randomization'' (FR), in which a random Gaussian deviation based on the
associated error bar, is then applied to each of the selected $N$
points.  This FR/RSS-altered subset of data points is then
cross-correlated as though it was real data.  The peak of the
cross-correlation function, $r_{\rm max}$, which occurs at a time lag
$\tau_{\rm peak}$ is determined, as is the centroid, $\tau_{\rm cent}$,
which is computed from those points near the peak with $r \geq 0.8
r_{\rm max}$.  A cross-correlation peak distribution (CCPD) for
$\tau_{\rm peak}$ and a cross-correlation centroid distribution (CCCD)
for $\tau_{\rm cent}$ is built with a large number ($N = 10000$) of
Monte Carlo realizations.  We take the average value of the CCPD to be
$\tau_{\rm peak}$ and the average value of the CCCD to be $\tau_{\rm
cent}$, with uncertainties $\Delta \tau_{\rm upper}$ and $\Delta
\tau_{\rm lower}$ such that 15.87\% of the CCCD realizations have values
$\tau > \tau_{\rm cent} + \Delta \tau_{\rm upper}$ and 15.87\% have
values $\tau < \tau_{\rm cent} - \Delta \tau_{\rm lower}$, with similar
uncertainties for $\tau_{\rm peak}$.  These definitions of the
uncertainty correspond to $\pm 1 \sigma$ for a Gaussian distribution.
We find an observed-frame lag time of $\tau_{\rm cent} =
6.59^{+1.12}_{-0.76}$~days and $\tau_{\rm peak} =
6.10^{+1.20}_{-0.60}$~days for NGC~4151.

\subsection{Line Width Measurement}

In order to calculate the black hole mass, we need to measure the width
of the H$\beta$ line in the variable (rms) part of the spectrum.  The
mean and rms spectra, shown in Fig.~2, were calculated using the full
set of MDM spectra.  To measure the line width, we first interpolate the
rms continuum underneath the H$\beta$ emission line by choosing
continuum windows on either side of the line.  The line width is then
characterized by its full-width at half maximum ($FWHM$) and by the line
dispersion (the second moment of the line profile) $\sigma_{\rm line}$,
as described by \citet{peterson04}.  In order to characterize the
uncertainties in each of these measurements, we use the following
procedure, as described by \citet{peterson04}.  For a set of $N$
spectra, we select $N$ spectra at random, without regard to whether a
spectrum has been previously selected or not.  The $N$ randomly selected
spectra are then used to construct mean and rms spectra from which the
line width measurements are made.  This process is one Monte-Carlo
realization, and a large number ($N = 10000$) of these realizations
gives a mean and standard deviation for each of the measurements of the
line width.  The H$\beta$ emission-line width measurements we determined
from the rms spectrum for NGC~4151 are $\sigma_{\rm line} = 2680 \pm
64$~km~s$^{-1}$ and $FWHM = 4711 \pm 750$~km~s$^{-1}$ in the restframe
of NGC~4151.

\subsection{Black Hole Mass}

The mass of the black hole is given by

\begin{equation}
M_{\rm BH} = \frac{f c \tau \Delta V^2}{G},
\end{equation}

\noindent where $\tau$ is the emission-line time delay, $\Delta V$ is the width
of the emission line, and $G$ is the gravitational constant.  The factor
$f$ depends on the geometry, kinematics, and inclination of the BLR and
has been shown by \citet{onken04} to have an average value of $\langle f
\rangle = 5.5$ when normalizing the AGN relationship between black hole
mass and stellar velocity dispersion ($M_{\rm BH} - \sigma_*$) to that
of quiescent galaxies.

To estimate the mass of NGC~4151, we use $\tau_{\rm cent}$ for the time
delay and $\sigma_{\rm line}$ for the line width of H$\beta$.  We find
$M_{\rm BH} = (5.1^{+0.9}_{-0.6}) \times 10^{7} M_{\odot}$.  We would,
however, like to remind the reader that all reverberation-based masses
are likely to be uncertain by a factor of 3-4.

\section{DISCUSSION AND CONCLUSIONS}

It has been clear for many years that the BLR in AGN is stratified in
ionization.  Emission lines arising from regions of more highly-ionized
gas have shorter time lags, and lines arising from regions of less
highly-ionized gas have longer time lags
(\citealt{clavel91,peterson91,dietrich93,maoz93}).  Additionally, it has
been known that lines arising from more highly-ionized gas have broader
profiles than lines arising from less highly-ionized gas
\citep{osterbrock82}.  If gravity dominates the kinematics of the BLR
gas, then a virial relationship between the time lag and the emission
line width, $\Delta V \propto \tau^{-1/2}$, will exist.  In fact, this
relationship has been found in NGC~5548 \citep{peterson99} as well as a
handful of other objects (\citealt{peterson00,onken02,kollatschny03}).

We can investigate the presence of the expected virial relationship
between time lag and line width in NGC~4151 by comparing our
measurements with those of other monitoring programs for NGC~4151.  A
reanalysis of previous ultraviolet monitoring campaigns for NGC~4151 was
recently undertaken by \citet*{metzroth06}.  Even though the {\it
International Ultraviolet Explorer} ({\it IUE}) data set of
\citet{clavel90} is undersampled, and the {\it IUE} data set of
\citet{ulrich96} is of a fairly short duration, the estimate of $M_{\rm
BH}$ determined by \citet{metzroth06} of $(4.14 \pm 0.73) \times 10^7
M_{\odot}$ is in good agreement with the value we obtain above.

Figure~3 shows the lag and line-width measurements (given by $\tau_{\rm
cent}$ and $\sigma_{\rm line}$) for NGC~4151 from \citet{metzroth06} and
from this work.  We fit a power-law to the UV data from
\citet{metzroth06} and the optical data from this work.  Even with the
small range of values available for $\tau$ and $\Delta V$, we find that
the best fit gives a slope of $\alpha = -0.76 \pm 0.24$, consistent with
the slope of $\alpha = -0.5$ expected if the virial relationship between
the size of the BLR and the velocity of the gas exists.  We do not
include measurements from two earlier ground-based monitoring campaigns
on NGC~4151, one in 1988 \citep{maoz91} and one in 1993--94
\citep{kaspi96}, for reasons we will briefly mention here, but are
described in detail in \citet{metzroth06}.  The \citeauthor{kaspi96}
data set has a rather unconstrained lag time that is consistent with
zero once the monotonic increase in the continuum and line flux is
removed.  The \citeauthor{maoz91} data set, on the other hand, has a
variable line spread function as well as low spectral resolution that
causes the [\ion{O}{3}] lines to blend together.  These factors make the
line width determinations suspect.

Although the line widths in the \citet{maoz91} campaign appear to be
unreliable, the time lag between the continuum and H$\beta$ line is well
defined.  The reanalysis of this data set in \citet{peterson04} gives a
lag time of $\tau_{\rm cent} = 11.5 \pm 3.7$~days, which is quite a bit
longer than the lag time found in this study.  This discrepancy
is not surprising given the variable nature of such objects.  For
comparison, the lag time of the H$\beta$ line with respect to the continuum
has been observed to range between 6.5 and 26.5~days in NGC~5548
 as the ionizing flux level changes over time \citep{peterson04}.  With
such a difference in the measured lag times of the H$\beta$ line in
NGC~4151, we would expect that the flux would be 30\% higher in the
\citeauthor{maoz91} campaign than in the current study, $F(H\beta) = (20.1
\pm 2.5) \times 10^{-13}$~ergs~s$^{-1}$~cm$^{-2}$ .  In actuality, the
flux during the \citeauthor{maoz91} campaign, $F(H\beta) = (21.5 \pm
3.4) \times 10^{-13}$~ergs~s$^{-1}$~cm$^{-2}$, is 1.5$\sigma$ below the
expected value.  However, this is not outside the observed scatter for
the H$\beta$ lag times and luminosities measured for NGC~5548 (see
Fig.~3 of \citealt{peterson02}).

Table~3 lists the black hole mass estimates for NGC~4151 calculated
using a variety of data sets and methods.  We show in bold face the
weighted mean of the black hole mass estimated using the H$\beta$ lag
and line-width and the masses determined with the UV emission lines
analyzed by \citet{metzroth06}, a value of $M_{\rm BH} =
4.57^{+0.57}_{-0.47} \times 10^7 M_{\odot}$, which we take to be the
best reverberation-mapping mass estimate for NGC~4151.  We can compare
the mass determined using reverberation-mapping to the mass expected
from the $M_{\rm BH} - \sigma_*$ relationship.  The stellar velocity
dispersion of NGC~4151 was measured by \citet{ferrarese01} to be
$\sigma_* = 93 \pm 14$ km s$^{-1}$.  Using the slope and zero-point
determined by \citet{tremaine02}, the $M_{\rm BH} - \sigma_*$
relationship predicts a black hole mass of $M_{\rm BH} =
6.2^{+5.9}_{-3.0} \times 10^6 M_{\odot}$, with both the scatter in the
$M_{\rm BH} - \sigma_*$ relationship and the uncertainty in the
measurement of $\sigma_*$ accounting for the large errors on the
predicted black hole mass.  This is about a factor of 7 smaller than the
reverberation measurement.  We caution, however, that the calibration of
the reverberation-based black hole mass scale \citep{onken04} is
averaged over many objects.  The scale factor $f$ in eq.\ (1) depends on
the structure, kinematics, and inclination of the BLR
(cf. \citealt{collin06}) and is expected to be different for each AGN.
Use of the mean scaling factor $\langle f \rangle = 5.5$ removes bias
from the sample, in the sense that as many masses are underestimated as
overestimated. Given that there is a factor of 3--4 scatter around the
normalized AGN $M_{\rm BH} - \sigma_*$ relationship, a single object
deviating by a factor of seven is hardly inconsistent.

In summary, the ground-based reverberation-mapping data presented here
for NGC~4151 supersedes previous optical data sets on this AGN.  We have
measured the lag of the H$\beta$ line to the continuum flux at 5100~\AA\
and found it to be $6.6^{+1.1}_{-0.8}$ days.  We have calculated the
weighted mean of black hole masses determined using H$\beta$ as well as
the UV emission lines analyzed by \citet{metzroth06}, which we find to
be $4.57^{+0.57}_{-0.47} \times 10^7 M_{\odot}$.  We also find that the
data are consistent with a virial relationship between the size of the
BLR and the velocity of BLR gas in NGC~4151.

\acknowledgements
We would like to thank an anonymous referee for comments that added to
the presentation of this paper.  We are grateful for support of this
work by the National Science Foundation through grant AST-0205964 to The
Ohio State University and the Civilian Research and Development
Foundation through grant UP1-2549-CR-03.  MCB is supported by a Graduate
Fellowship of the National Science Foundation.  KDD is supported by a
GK-12 Fellowship of the National Science Foundation.  EMC gratefully
acknowledges support from PPARC.  MV acknowledges financial support from
NSF grant AST-0307384 to the University of Arizona.  This research has
made use of the NASA/IPAC Extragalactic Database (NED) which is operated
by the Jet Propulsion Laboratory, California Institute of Technology,
under contract with the National Aeronautics and Space Administration.



\clearpage

\begin{deluxetable}{cccc}
\tablecolumns{4}
\tablewidth{0pt}
\tablecaption{Continuum and H$\beta$ Fluxes for NGC~4151}
\tablehead{
\colhead{JD\tablenotemark{a}} &
\colhead{F$_{\lambda}$ (5100 \AA)} &
\colhead{H$\beta$ $\lambda 4681$} &
\colhead{Observatory}\\
\colhead{(-2,450,000)} &
\colhead{($10^{-15}$ ergs s$^{-1}$ cm$^{-2}$ \AA$^{-1}$)} &
\colhead{($10^{-13}$ ergs s$^{-1}$ cm$^{-2}$)} &
\colhead{Code\tablenotemark{b}}}

\startdata

3430.020 & 26.96 $\pm$	0.62	& 21.40 $\pm$	0.27 &	M \\
3430.090 & 26.12 $\pm$	0.43	& \nodata	     &  H \\
3430.785 & 26.03 $\pm$	0.60	& 21.13 $\pm$	0.27 &	M \\
3431.762 & 27.13 $\pm$	0.63	& 21.90 $\pm$	0.28 &	M \\
3432.773 & 28.24 $\pm$	0.65	& 22.42 $\pm$	0.28 &	M \\
3432.809 & 28.10 $\pm$	0.54	& \nodata	     &  H \\
3433.750 & 27.18 $\pm$	0.63	& 21.94 $\pm$	0.28 &	M \\
3434.941 & 28.55 $\pm$	0.63	& \nodata	     &  H \\
3436.910 & 29.87 $\pm$	0.46	& \nodata	     &  H \\
3437.336 & 31.62 $\pm$	0.70	& 22.41 $\pm$	0.40 &	C \\
3437.812 & 28.14 $\pm$	0.65	& 23.33 $\pm$	0.29 &	M \\
3438.148 & 29.00 $\pm$	0.27	& \nodata	     &  H \\
3438.750 & 28.35 $\pm$	0.66	& 23.69 $\pm$	0.30 &	M \\
3439.723 & 27.10 $\pm$	0.63	& 23.37 $\pm$	0.29 &	M \\
3440.754 & 26.03 $\pm$	0.60	& 23.28 $\pm$	0.29 &	M \\
3441.898 & 24.89 $\pm$	0.58	& 23.88 $\pm$	0.30 &	M \\
3442.754 & 23.53 $\pm$	0.54	& 23.38 $\pm$	0.30 &	M \\
3442.820 & 24.15 $\pm$	1.15	& \nodata	     &  H \\
3443.754 & 22.38 $\pm$	0.52	& 23.24 $\pm$	0.29 &	M \\
3444.395 & 22.90 $\pm$	0.51	& 23.08 $\pm$	0.42 &	C \\
3445.391 & 21.43 $\pm$	0.47	& 22.69 $\pm$	0.41 &	C \\
3445.766 & 21.49 $\pm$	0.50	& 22.36 $\pm$	0.28 &	M \\
3446.332 & 21.28 $\pm$	0.47	& 21.56 $\pm$	0.39 &	C \\
3446.746 & 21.96 $\pm$	0.51	& 21.85 $\pm$	0.28 &	M \\
3447.828 & 21.09 $\pm$	0.69	& \nodata	     &  H \\
3447.930 & 21.66 $\pm$	0.50	& 20.83 $\pm$	0.26 &	M \\
3450.785 & 21.64 $\pm$	0.50	& 19.76 $\pm$	0.25 &	M \\
3451.777 & 21.62 $\pm$	0.50	& 19.34 $\pm$	0.24 &	M \\
3451.988 & 22.14 $\pm$	0.16	& \nodata	     &  H \\
3452.695 & 21.65 $\pm$	0.50	& 18.99 $\pm$	0.24 &	M \\
3456.766 & 22.07 $\pm$	0.51	& 19.01 $\pm$	0.24 &	M \\
3458.910 & 21.90 $\pm$	0.51	& 17.89 $\pm$	0.23 &	M \\
3459.773 & 21.25 $\pm$	0.49	& 18.10 $\pm$	0.23 &	M \\
3460.863 & 21.74 $\pm$	0.50	& 18.00 $\pm$	0.23 &	M \\
3461.746 & 21.67 $\pm$	0.50	& 17.86 $\pm$	0.23 &	M \\
3462.781 & 21.29 $\pm$	0.49	& 17.61 $\pm$	0.22 &	M \\
3463.367 & 23.37 $\pm$	0.52	& 17.74 $\pm$	0.32 &	C \\
3464.371 & 23.02 $\pm$	0.51	& 17.52 $\pm$	0.32 &	C \\
3464.809 & 22.27 $\pm$	0.52	& 18.00 $\pm$	0.23 &	M \\
3465.805 & 21.93 $\pm$	0.51	& 17.52 $\pm$	0.22 &	M \\
3466.793 & 22.68 $\pm$	0.52	& 17.29 $\pm$	0.22 &	M \\
3466.941 & 21.87 $\pm$	0.16	& \nodata	     &  H \\
3467.895 & 22.03 $\pm$	0.51	& 17.55 $\pm$	0.22 &	M \\
3468.781 & 20.66 $\pm$	0.48	& 17.70 $\pm$	0.22 &	M \\
3469.398 & 21.01 $\pm$	0.46	& 17.46 $\pm$	0.31 &	C \\
3469.734 & 20.80 $\pm$	0.48	& 17.56 $\pm$	0.22 &	M \\
3469.941 & 20.64 $\pm$	0.11	& \nodata	     &  H \\
3470.312 & 20.80 $\pm$	0.46	& 17.28 $\pm$	0.31 &	C \\
3470.785 & 20.30 $\pm$	0.47	& 17.46 $\pm$	0.22 &	M \\
3471.785 & 20.66 $\pm$	0.48	& 16.90 $\pm$	0.21 &	M \\

\enddata

\tablenotetext{a}{The Julian Date listed is the midpoint of the observation}.

\tablenotetext{b}{Observatory Codes:
		  {\bf C} = CrAO 2.6-m Shajn Telescope + Nasmith 
				Spectrograph;
		  {\bf H} = Haleakala Observatories 2.0-m MAGNUM Telescope 
				+ MIP;
		  {\bf M} = MDM Observatory 1.3-m McGraw-Hill Telescope + 
				CCDS.}

\end{deluxetable}

\begin{deluxetable}{lccccccc}
\tablecolumns{8}
\tablewidth{0pt}
\tablecaption{Light Curve Statistics}
\tablehead{
\colhead{} &
\colhead{} &
\multicolumn{2}{c}{Sampling} &
\colhead{} &
\colhead{Mean} &
\colhead{} &
\colhead{} \\
\colhead{Time} &
\colhead{} &
\multicolumn{2}{c}{Interval (days)} &
\colhead{Mean} &
\colhead{Fractional} &
\colhead{} &
\colhead{} \\
\colhead{Series} &
\colhead{$N$} &
\colhead{$\langle T \rangle$} &
\colhead{$T_{\rm median}$} &
\colhead{Flux\tablenotemark{a}} &
\colhead{Error} &
\colhead{$F_{\rm var}$} &
\colhead{$R_{\rm max}$}\\
\colhead{(1)} &
\colhead{(2)} &
\colhead{(3)} &
\colhead{(4)} &
\colhead{(5)} &
\colhead{(6)} &
\colhead{(7)} &
\colhead{(8)}}
\startdata

5100 \AA & 37 & 1.2 & 1.0 & $23.8 \pm 3.0$ & 0.019 & 0.124 & $1.45 \pm 0.03$ \\ 
H$\beta$ & 34 & 1.3 & 1.0 & $20.1 \pm 2.5$ & 0.012 & 0.121 & $1.41 \pm 0.02$ \\

\enddata
\tablenotetext{a}{Fluxes are in the same units as those used in Table~1.}
\end{deluxetable}

\begin{deluxetable}{ccc}
\tablecolumns{3}
\tablewidth{0pt}
\tablecaption{Black Hole Mass Estimates}
\tablehead{
\colhead{Description} &
\colhead{$M_{\rm BH}$ ($10^7 M_{\odot}$)} &
\colhead{References}}
\startdata

Previous H$\alpha$ + H$\beta$		& $1.33^{+0.46}_{-0.46}$ & 1 \\
UV Weighted Mean			& $4.14^{+0.73}_{-0.73}$ & 2 \\
New H$\beta$				& $5.05^{+0.89}_{-0.63}$ & 3 \\
{\bf UV + H$\beta$ Weighted Mean} & {\boldmath $4.57^{+0.57}_{-0.47}$} & 3 \\
$M_{\rm BH} - \sigma_*$ Estimate	& $0.62^{+0.59}_{-0.30}$ & 4,5 \\

\enddata
\tablerefs{(1) \citet{peterson04};
	   (2) \citet{metzroth06};
	   (3) this work;
	   (4) \citet{ferrarese01};
	   (5) \citet{tremaine02}.}
\end{deluxetable}

\clearpage

\begin{figure}
\figurenum{1}
\epsscale{1}
\plotone{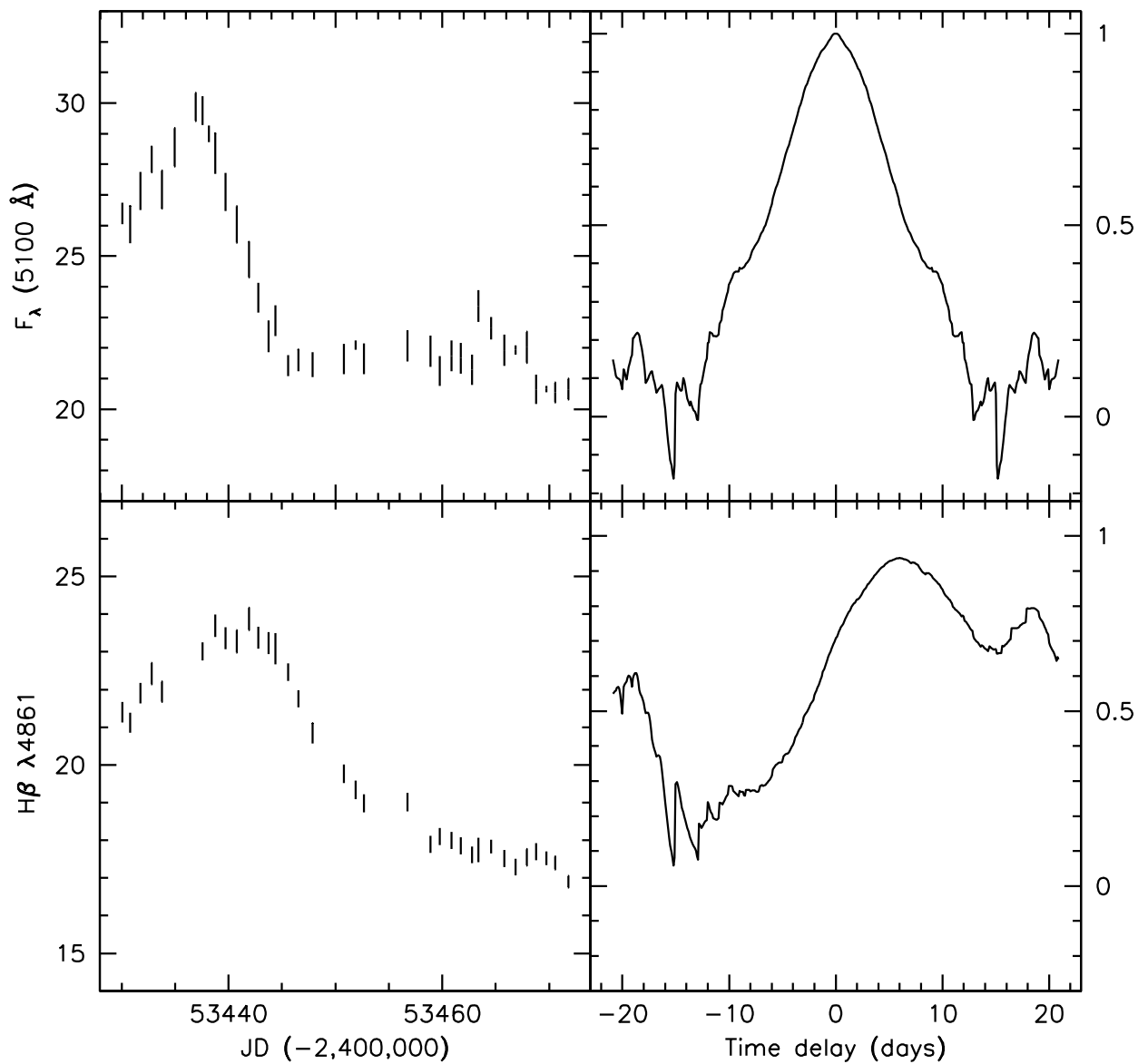}
\caption{The left panel shows the lightcurves for the continuum region 
         at 5100 \AA\ and for the H$\beta$ $\lambda 4861$ line.
         Measurements taken within a 0.5 day bin have been averaged
         together.  The continuum flux is in units of $10^{-15}$ erg
         s$^{-1}$ cm$^{-2}$ \AA$^{-1}$, and the H$\beta$ flux is in
         units of $10^{-13}$ erg s$^{-1}$ cm$^{-2}$ \AA$^{-1}$.  The
         right panel shows the result of cross-correlation with the
         continuum light curve; the top right panel is therefore the
         continuum autocorrelation function.}
\end{figure}

\clearpage

\begin{figure}
\figurenum{2}
\epsscale{1}
\plotone{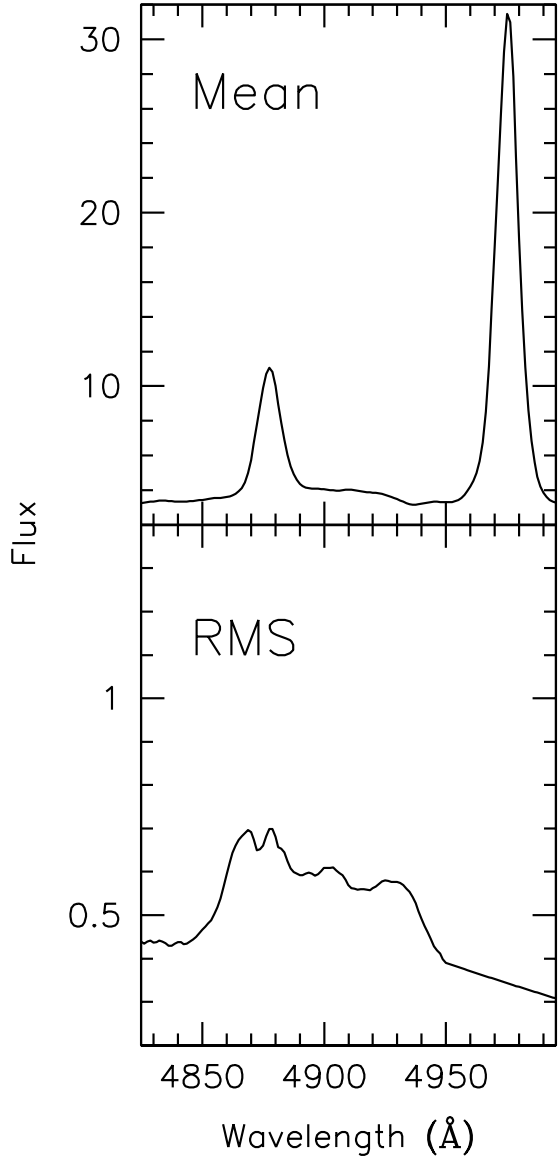}
\caption{Mean and RMS of the MDM spectra in the observed frame of NGC~4151.
	 The narrow components of the lines have been subtracted from
	 each spectrum before creating the RMS spectrum to avoid the
	 residuals of the very strong [\ion{O}{3}] lines in NGC~4151.
	 Some of the structure along the top of the H$\beta$ broad line
	 is likely due to the imperfect subtraction of the narrow
	 component of the line.  The flux for both spectra is in units
	 of $10^{-14}$ erg s$^{-1}$ cm$^{-2}$ \AA$^{-1}$.}

\end{figure}

\clearpage

\begin{figure}
\figurenum{3}
\epsscale{1}
\plotone{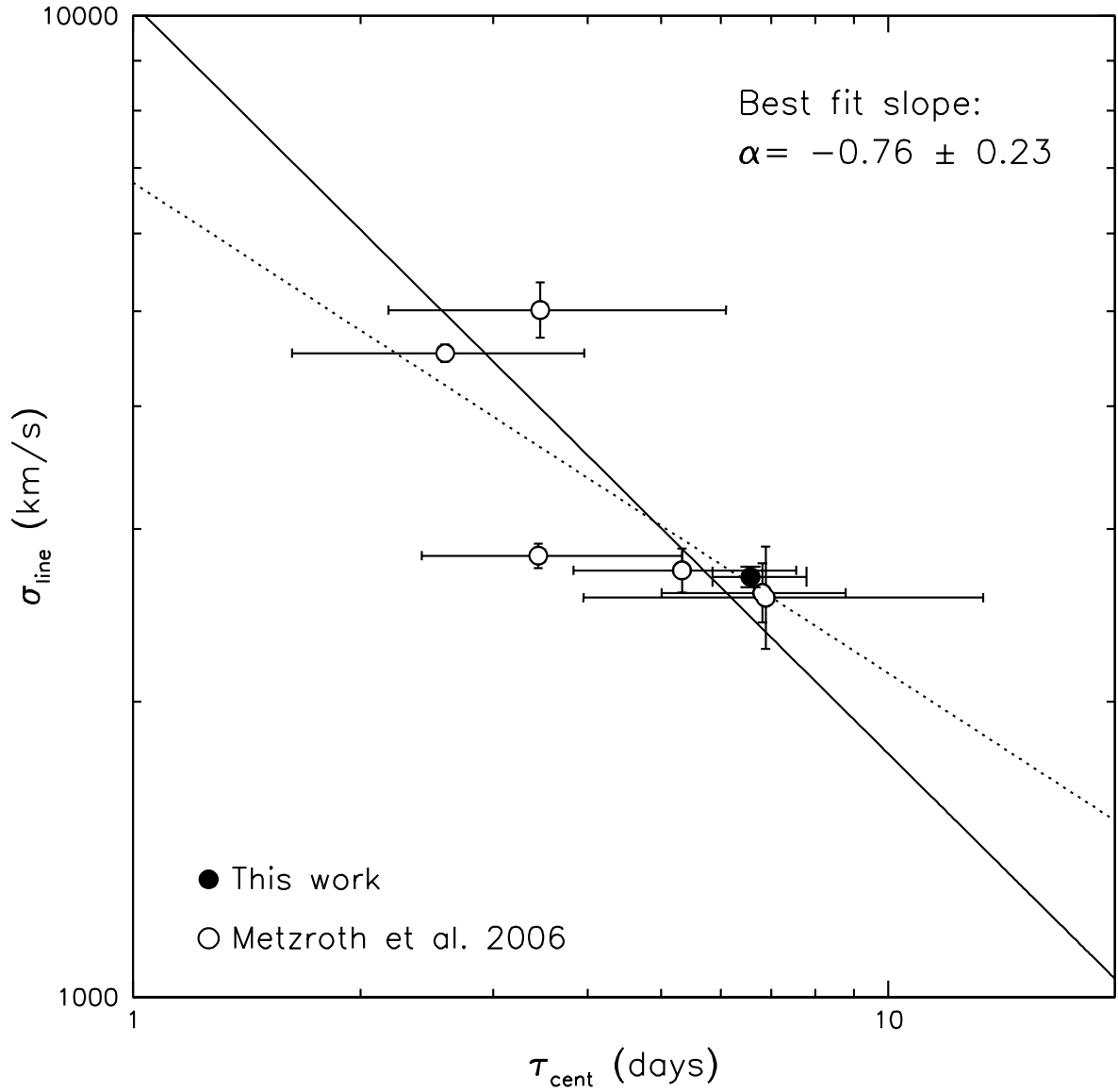}
\caption{Rest-frame emission-line widths versus time lags for lines in 
	 NGC~4151.  The filled circle is the data from this work.  The
	 open circles are UV emission lines from \citet{metzroth06}.
	 The solid line has a slope of $-0.76 \pm 0.24$ and is the best
	 fit to the circles.  The dotted line has a forced slope of
	 $-0.5$, i.e. a virial relationship.}
\end{figure}


\begin{thebibliography}{42}
\expandafter\ifx\csname natexlab\endcsname\relax\def\natexlab#1{#1}\fi

\bibitem[{{Antonucci} \& {Cohen}(1983)}]{antonucci83}
{Antonucci}, R.~R.~J., \& {Cohen}, R.~D. 1983, \apj, 271, 564

\bibitem[{{Bentz} {et~al.}(2006){Bentz}, {Peterson}, {Pogge}, {Vestergaard}, \&
  {Onken}}]{bentz06}
{Bentz}, M.~C., {Peterson}, B.~M., {Pogge}, R.~W., {Vestergaard}, M., \&
  {Onken}, C.~A. 2006, \apj, in press (astro-ph/0602412)

\bibitem[{{Berezin} {et~al.}(1991){Berezin}, {Zuev}, {Kiryan}, {Rybakov},
  {Khvilivitskii}, {Ilin}, {Petrov}, {Savanov}, \& {Shcherbakov}}]{berezin91}
{Berezin}, V.~Y., {Zuev}, A.~G., {Kiryan}, G.~V., {Rybakov}, M.~I.,
  {Khvilivitskii}, A.~T., {Ilin}, I.~V., {Petrov}, P.~P., {Savanov}, I.~S., \&
  {Shcherbakov}, A.~G. 1991, Soviet Astronomy Letters, 17, 405

\bibitem[{{Blandford} \& {McKee}(1982)}]{blandford82}
{Blandford}, R.~D., \& {McKee}, C.~F. 1982, \apj, 255, 419

\bibitem[{{Clavel} {et~al.}(1990)}]{clavel90}
{Clavel}, J., {et~al.} 1990, \mnras, 246, 668

\bibitem[{{Clavel} {et~al.}(1991)}]{clavel91}
---. 1991, \apj, 366, 64

\bibitem[{{Collin} {et~al.}(2006){Collin}, {Kawaguchi}, {Peterson}, \&
  {Vestergaard}}]{collin06}
{Collin}, S., {Kawaguchi}, T., {Peterson}, B.~M., \& {Vestergaard}, M. 2006,
  \aap, in press (astro-ph/0603460)

\bibitem[{{Dietrich} {et~al.}(1993)}]{dietrich93}
{Dietrich}, M., {et~al.} 1993, \apj, 408, 416

\bibitem[{{Ferrarese} {et~al.}(2001){Ferrarese}, {Pogge}, {Peterson},
  {Merritt}, {Wandel}, \& {Joseph}}]{ferrarese01}
{Ferrarese}, L., {Pogge}, R.~W., {Peterson}, B.~M., {Merritt}, D., {Wandel},
  A., \& {Joseph}, C.~L. 2001, \apjl, 555, L79

\bibitem[{{Hunt} {et~al.}(1998){Hunt}, {Mannucci}, {Testi}, {Migliorini},
  {Stanga}, {Baffa}, {Lisi}, \& {Vanzi}}]{hunt98}
{Hunt}, L.~K., {Mannucci}, F., {Testi}, L., {Migliorini}, S., {Stanga}, R.~M.,
  {Baffa}, C., {Lisi}, F., \& {Vanzi}, L. 1998, \aj, 115, 2594

\bibitem[{{Kaspi} {et~al.}(1996)}]{kaspi96}
{Kaspi}, S., {et~al.} 1996, \apj, 470, 336

\bibitem[{{Kobayashi} {et~al.}(1998{\natexlab{a}}){Kobayashi}, {Yoshii},
  {Peterson}, {Minezaki}, {Enya}, {Suganuma}, \& {Yamamuro}}]{kobayashi98a}
{Kobayashi}, Y., {Yoshii}, Y., {Peterson}, B.~A., {Minezaki}, T., {Enya}, K.,
  {Suganuma}, M., \& {Yamamuro}, T. 1998{\natexlab{a}}, in Proc. SPIE, Vol.
  3354, 769--776

\bibitem[{{Kobayashi} {et~al.}(1998{\natexlab{b}})}]{kobayashi98b}
{Kobayashi}, Y., {et~al.} 1998{\natexlab{b}}, in Proc. SPIE, Vol. 3352,
  120--128

\bibitem[{{Kollatschny}(2003)}]{kollatschny03}
{Kollatschny}, W. 2003, \aap, 407, 461

\bibitem[{{Kollmeier} {et~al.}(2006)}]{kollmeier06}
{Kollmeier}, J.~A., {et~al.} 2006, \apj, in press (astro-ph/0508657)

\bibitem[{{Landolt}(1992)}]{landolt92}
{Landolt}, A.~U. 1992, \aj, 104, 340

\bibitem[{{Maoz} {et~al.}(1991)}]{maoz91}
{Maoz}, D., {et~al.} 1991, \apj, 367, 493

\bibitem[{{Maoz} {et~al.}(1993)}]{maoz93}
---. 1993, \apj, 404, 576

\bibitem[{{McLure} \& {Jarvis}(2002)}]{mclure02}
{McLure}, R.~J., \& {Jarvis}, M.~J. 2002, \mnras, 337, 109

\bibitem[{{Metzroth} {et~al.}(2006){Metzroth}, {Onken}, \&
  {Peterson}}]{metzroth06}
{Metzroth}, K.~G., {Onken}, C.~A., \& {Peterson}, B.~M. 2006, \apj, in press

\bibitem[{{Minezaki} {et~al.}(2004){Minezaki}, {Yoshii}, {Kobayashi}, {Enya},
  {Suganuma}, {Tomita}, {Aoki}, \& {Peterson}}]{minezaki04}
{Minezaki}, T., {Yoshii}, Y., {Kobayashi}, Y., {Enya}, K., {Suganuma}, M.,
  {Tomita}, H., {Aoki}, T., \& {Peterson}, B.~A. 2004, \apjl, 600, L35

\bibitem[{{Onken} {et~al.}(2004){Onken}, {Ferrarese}, {Merritt}, {Peterson},
  {Pogge}, {Vestergaard}, \& {Wandel}}]{onken04}
{Onken}, C.~A., {Ferrarese}, L., {Merritt}, D., {Peterson}, B.~M., {Pogge},
  R.~W., {Vestergaard}, M., \& {Wandel}, A. 2004, \apj, 615, 645

\bibitem[{{Onken} \& {Peterson}(2002)}]{onken02}
{Onken}, C.~A., \& {Peterson}, B.~M. 2002, \apj, 572, 746

\bibitem[{{Osterbrock} \& {Shuder}(1982)}]{osterbrock82}
{Osterbrock}, D.~E., \& {Shuder}, J.~M. 1982, \apjs, 49, 149

\bibitem[{{Peterson}(1993)}]{peterson93}
{Peterson}, B.~M. 1993, \pasp, 105, 247

\bibitem[{{Peterson} \& {Wandel}(1999)}]{peterson99}
{Peterson}, B.~M., \& {Wandel}, A. 1999, \apjl, 521, L95

\bibitem[{{Peterson} {et~al.}(1998){Peterson}, {Wanders}, {Bertram}, {Hunley},
  {Pogge}, \& {Wagner}}]{peterson98}
{Peterson}, B.~M., {Wanders}, I., {Bertram}, R., {Hunley}, J.~F., {Pogge},
  R.~W., \& {Wagner}, R.~M. 1998, \apj, 501, 82

\bibitem[{{Peterson} {et~al.}(1991)}]{peterson91}
{Peterson}, B.~M., {et~al.} 1991, \apj, 368, 119

\bibitem[{{Peterson} {et~al.}(2000)}]{peterson00}
---. 2000, \apj, 542, 161

\bibitem[{{Peterson} {et~al.}(2002)}]{peterson02}
---. 2002, \apj, 581, 197

\bibitem[{{Peterson} {et~al.}(2004)}]{peterson04}
---. 2004, \apj, 613, 682

\bibitem[{{Peterson} {et~al.}(2005)}]{peterson05}
---. 2005, \apj, 632, 799

\bibitem[{{Tremaine} {et~al.}(2002)}]{tremaine02}
{Tremaine}, S., {et~al.} 2002, \apj, 574, 740

\bibitem[{{Ulrich} \& {Horne}(1996)}]{ulrich96}
{Ulrich}, M.-H., \& {Horne}, K. 1996, \mnras, 283, 748

\bibitem[{{van Groningen} \& {Wanders}(1992)}]{vanGroningen92}
{van Groningen}, E., \& {Wanders}, I. 1992, \pasp, 104, 700

\bibitem[{{Vestergaard}(2002)}]{vestergaard02}
{Vestergaard}, M. 2002, \apj, 571, 733

\bibitem[{{Vestergaard}(2004)}]{vestergaard04}
---. 2004, \apj, 601, 676

\bibitem[{{Vestergaard} \& {Peterson}(2006)}]{vestergaard06}
{Vestergaard}, M., \& {Peterson}, B.~M. 2006, \apj, 641, 689

\bibitem[{{Wandel} {et~al.}(1999){Wandel}, {Peterson}, \& {Malkan}}]{wandel99}
{Wandel}, A., {Peterson}, B.~M., \& {Malkan}, M.~A. 1999, \apj, 526, 579

\bibitem[{{White} \& {Peterson}(1994)}]{white94}
{White}, R.~J., \& {Peterson}, B.~M. 1994, \pasp, 106, 879

\bibitem[{{Yoshii}(2002)}]{yoshii02}
{Yoshii}, Y. 2002, in New Trends in Theoretical and Observational Cosmology,
  ed. K.~Sato \& T.~Shiromizu (Tokyo: Universal Academy), 235

\bibitem[{{Yoshii} {et~al.}(2003){Yoshii}, {Kobayashi}, \&
  {Minezaki}}]{yoshii03}
{Yoshii}, Y., {Kobayashi}, Y., \& {Minezaki}, T. 2003, \baas, 202, 3803

\end{thebibliography}
\end{document}